\documentclass[apj]{emulateapj}
\usepackage{epsfig}

\def\ts{{\thinspace}}
\def\simgt{\lower.5ex\hbox{$\; \buildrel > \over \sim \;$}}
\def\simlt{\lower.5ex\hbox{$\; \buildrel < \over \sim \;$}}

\def\s{\phantom}

\def\etal{{et~al.}}
\def\amin{\ifmmode^{\prime}\else$^{\prime}$\fi}
\def\asec{\ifmmode^{\prime\prime}\else$^{\prime\prime}$\fi}
\def\ts{{\thinspace}}
\def\simgt{\lower.5ex\hbox{$\; \buildrel > \over \sim \;$}}
\def\simlt{\lower.5ex\hbox{$\; \buildrel < \over \sim \;$}}

\newcommand\asca{{\it ASCA\/}}

\newcommand\chandra{{\it Chandra}}

\newcommand\hess{{\it HESS\/}}

\newcommand\snr{G12.82$-$0.02}
\newcommand\tev{\hbox{HESS~J1813$-$178}}
\newcommand\psr{CXOU J181335.1$-$174957}

\def\ref#1{{\par\noindent \hangindent=1em\hangafter=1
 #1.\par}}

\slugcomment{To Appear in the Astrophysical Journal}

\shorttitle{Putative Pulsar Coincident with \tev}
\shortauthors{Helfand \etal}

\begin{document}

\title{Discovery of the Putative Pulsar and Wind Nebula Associated
with the TeV Gamma-ray Source \tev} 

\author{ D.~J.~Helfand, E.~V.~Gotthelf, J.~P.~Halpern, F.~Camilo, D.~R.~Semler}

\affil{Columbia Astrophysics Laboratory, Columbia University, 550 West
120$^{th}$ Street, New York, NY 10027, USA}

\author{R.~H.~Becker}

\affil{Department of Physics, University of California, Davis, 1 Shields Avenue, Davis, CA 95616; and Institute of Geophysics and Planetary Physics, Lawrence Livermore National Laboratory, 7000 East Avenue, Livermore, CA 94550}

\and

\author{R.~L.~White}

\affil{Space Telescope Science Institute, 3700 San Martin Drive, Baltimore, MD 21218}

\begin{abstract}

We present a \chandra\ X-ray observation of \snr, a shell-like radio
supernova remnant coincident with the TeV gamma-ray source \tev. We
resolve the X-ray emission from the co-located \asca\ source into a
point source surrounded by structured diffuse emission that fills the
interior of the radio shell. The morphology of the diffuse emission
strongly resembles that of a pulsar wind nebula. The spectrum of the
compact source is well-characterized by a power-law with index $\Gamma
\approx 1.3$, typical of young and energetic rotation-powered pulsars.
For a distance of 4.5~kpc, consistent with the X-ray absorption and an
association with the nearby star formation region W33, the $2-10$~keV
X-ray luminosities of the putative pulsar and nebula are $L_{PSR} =
3.2 \times 10^{33}$~ergs~s$^{-1}$ and $L_{PWN} = 1.4 \times
10^{34}$~ergs~s$^{-1}$, respectively. Both the flux ratio of
$L_{PWN}/L_{PSR} = 4.3$ and the total luminosity of this system
predict a pulsar spin-down power of $\dot E > 10^{37}$~ergs~s$^{-1}$,
placing it within the ten most energetic young pulsars in the Galaxy.
A deep search for radio pulsations using the Parkes telescope sets an
upper-limit of $\approx 0.07$~mJy at 1.4~GHz for periods $\simgt
50$~ms. We discuss the energetics of this source, and consider
briefly the proximity of bright \ion{H}{2} regions to this and several
other \hess\ sources, which may produce their TeV emission via inverse
Compton scattering.

\end{abstract}

\section{Introduction}

The HESS observatory has revolutionized the field of TeV gamma-ray
astronomy, opening a new window onto the highest energy processes
occurring in our Galaxy and beyond. Of the 21 Galactic TeV sources
detected by \hess\ during the first two years of four-telescope
operation, firm identifications have been made for only seven objects
(\hess\ ``A'' class sources -- Funk et al. 2006a). Of these sources,
nearly all are associated with supernova products: four with bright
pulsar wind nebula (PWNe -- three of which contain detected young,
energetic pulsars), and two with non-thermal shell-type supernova
remnants (SNRs); the remaining object is associated with a high-mass
X-ray binary system. There are five less secure associations with
SNRs/PWNe (\hess\ ``B/C'' class sources). The remaining nine
\hess\ sources have yet to be identified with a known object at any
other wavelength.

An opportunity to study the origin of SNR TeV emission is provided by
the coincidence of the unidentified TeV source \tev\ with a previously
uncatalogued shell-type radio supernova remnant \snr\ (Brogan et al.
2005; Ubertini et al. 2005; Helfand et al. 2005).  This
low-surface-brightness, small-diameter ($\sim 2^{\prime}$) remnant
lies within the $1\sigma$ extent of \tev\ and is coincident with a
bright \asca\ X-ray source.  The possibility that this source
represented a third example of a shell-type SNR producing non-thermal
X-rays along with TeV gamma rays raised considerable interest;
however, as noted by the above authors, the source of the high-energy
emission could also be an energetic pulsar associated with this
apparently young remnant. Helfand \etal\ (2005) noted the proximity of
the star forming region W33 as a source of ambient photons for
producing gamma-rays from inverse Compton scattering.

We report here on a \chandra\ high-resolution image of \snr. The X-ray
flux from the remnant is resolved into diffuse non-thermal emission
that fills the interior of the radio SNR, and surrounds a bright
non-thermal point source. In \S2 we describe the \chandra\ observation
 of \snr, as well as supporting data we have collected in
other wavelength regimes. In \S3, we discuss our interpretation of
this system as a young, energetic rotation-powered pulsar associated
with the radio remnant and powering a pulsar wind
nebula\footnote{While this paper was in preparation, Funk et
al. (2006) submitted and posted a discussion of their {\it XMM\/}
observations of HESS J1813$-$178 that reaches similar
conclusions.}. We also consider the origin of TeV emission from this
and other HESS sources, suggesting that the presence or absence of a
high-intensity, local source of optical/IR photons may often be determinative
in the production of a TeV source. In the following text we assume a
distance to \snr\ of 4.5~kpc, as discussed in \S3.1.

\section{Observations and Results}

\subsection{The \chandra\ X-ray Observation}

A 30~ks X-ray observation of SNR \snr\ was obtained on 2006 September 15 UT
using the \chandra\ X-ray Observatory.  Data were collected with the
Advanced CCD Imaging Spectrometer (ACIS) in the focal plane, operating
in the nominal full-frame TIMED/VFAINT exposure mode. This detector is
sensitive to X-rays in the 0.3--12.0~keV energy range with a
resolution of $\Delta E /E \sim 0.06$ FWHM at 1~keV. The imaging
system offers an on-axis spatial resolution of $\sim 0\farcs5$, which
is also the instrument pixel size. The \asca\ source associated with
\snr\ was positioned on the ACIS-I3 front-illuminated CCD and offset
by $2^{\prime}$ from the nominal aimpoint to allow any extended X-ray
emission from the source to fall wholly on this one CCD. A total of
29.6~ks of live-time was accumulated with a CCD frame time of 3.241~s
(given the 1.3\% readout deadtime). No time filtering was necessary as
the background rate was stable over the course of the observation.
With a maximum count rate in a pixel of $< 0.004$~s$^{-1}$, photon
pile-up can be safely ignored.  We used the standard processed and
filtered event data, with the exception that the level~2 event file
was reprocessed to remove the pixel randomization, restoring a
slightly sharper image.  All data reduction and analysis was performed
using the CIAO (V3.3.0), FTOOLS (V6.0.4), CALDB 3.2.3, and XSPEC
(V12.2.1) X-ray analysis software packages.

%-----------------------------Figure Start--------------------------------
\begin{figure}
\centerline{
\hfill
\includegraphics[width=0.9\linewidth,angle=0,clip=true]{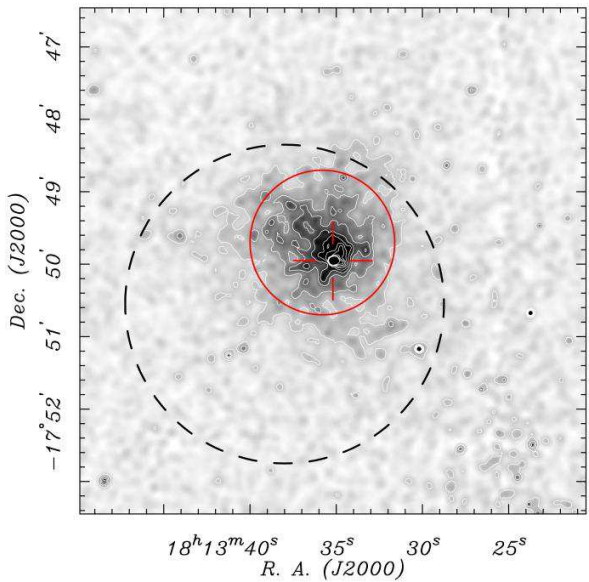}
\hfill
}
\caption{The 30~ks \chandra\ ACIS-I broad-band X-ray image of
\snr\ smoothed and scaled to highlight the extended diffuse 
emission. The cross marks the location of the point source.
The solid red circle ($r = 1^{\prime}$) gives the mean size of the 
supernova remnant's radio shell and the dashed circle
shows the $1\sigma$ extent of\ tev\ ($r = 2\farcm2$).
This portion of the image contains 23 other source significant at the 
$\geq 3\sigma$ level (see  Table~1), and
represents $17\%$ of the full ACIS-I field-of-view.
}
\label{fig1}
\end{figure}
%-----------------------------Figure End----------------------------------

%-----------------------------Figure Start--------------------------------
\begin{figure*}
\centerline{
\hfill
\includegraphics[width=0.45\linewidth,angle=0,clip=true]{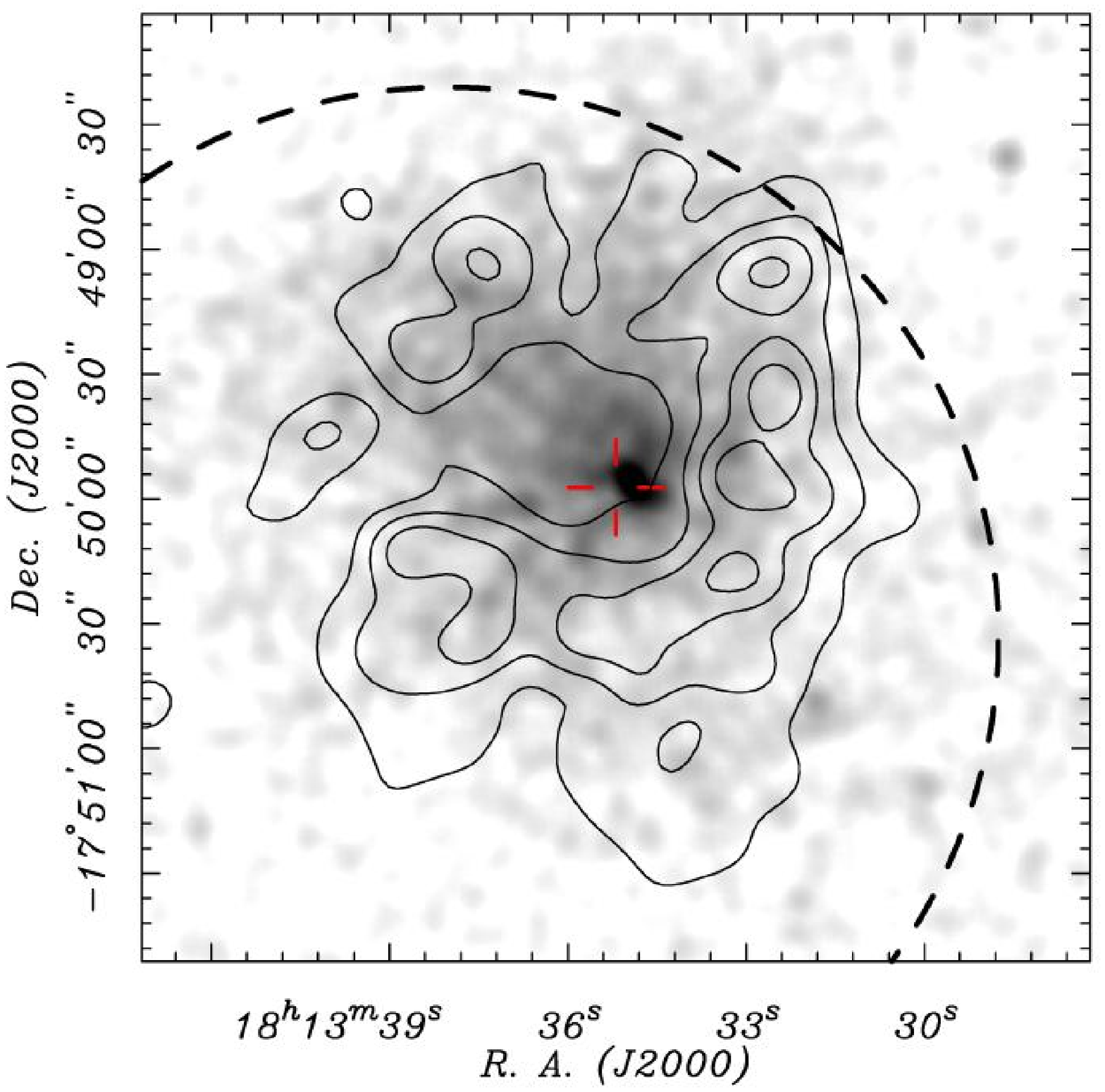}
\hfill
\includegraphics[width=0.45\linewidth,angle=0,clip=true]{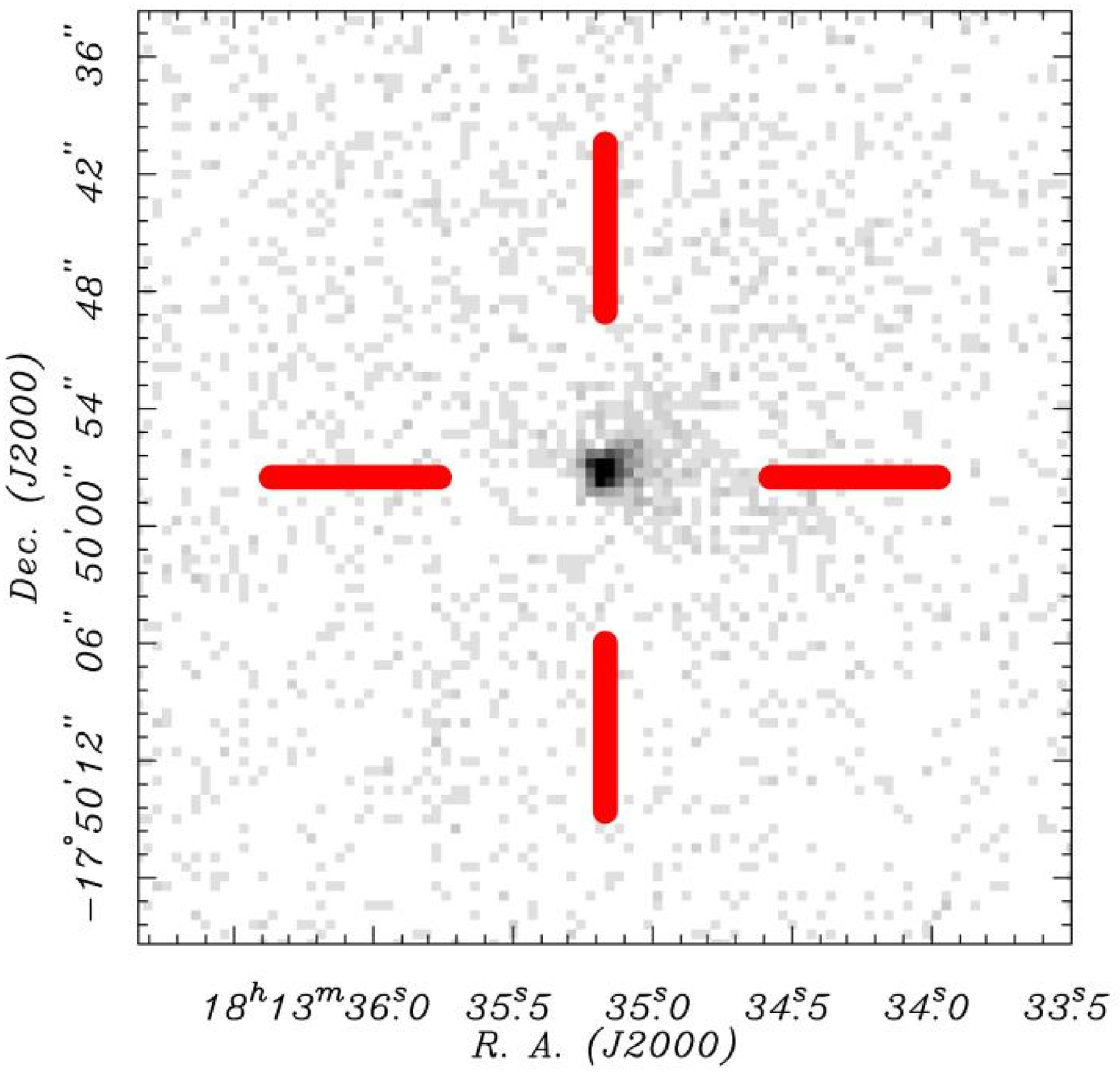}
\hfill
}
\caption{ The \chandra\ resolved X-ray emission component from \snr.
{\it Left\/}: The broad-band, smoothed X-ray image of \snr\ with VLA
radio contours overlaid; the point source, whose location is indicated
by the red cross, was removed to highlight the diffuse emission.  The
dashed circle illustrates the $1\sigma$ extent of \tev.  {\it
Right\/}: Zoom-in at full resolution of the X-ray image centered on
the point source, a candidate pulsar.  The intensity is scaled to
highlight the point source and inner filamentary feature. The red
cross is the same size (in arcseconds) and location in both panels.  }
\label{fig2}
\end{figure*}
%-----------------------------Figure End----------------------------------

As shown in Figures~1~and~2, the \chandra\ image of \snr\ resolves the
\asca\ source into diffuse X-ray emission that mostly fills the
radio shell and peaks toward a bright unresolved point source, the
brightest within the full $16^{\prime} \times 16^{\prime}$ ACIS-I
4-CCD field-of-view. This point source is offset from the
geometric center of the radio remnant by $\approx 20^{\prime\prime}$.
Near this source is a localized, oval-shaped structure extending to the
west/southwest (see Figure~2).  It is clear from a radial
profile centered on the point emission (Figure~3) that this excess
emission is prominent out to $5^{\prime\prime}$ and disappears into
the background at about $10^{\prime\prime}$.  The radial profile is
otherwise consistent with that of a point source when the large-scale
diffuse emission filling the radio shell is taken into account. The
overall morphology is highly reminiscent of a young, energetic pulsar
within a structured pulsar wind nebula.

The highest radio surface brightness region of the shell lies along
its western boundary, suggesting it has encountered a higher ambient
density in this direction. Taking the possibility of asymmetric
expansion into account, the location of the X-ray point source is
consistent with the location of the supernova explosion.
The offset of the point source from the center of the shell represents
a distance of only $\approx 0.4$~pc at the adopted distance of 4.5~kpc
(see \S3.1), implying $v\simlt 385$ km s$^{-1}$ if the remnant is
$\simgt 10^3$~yr old.

%-----------------------------Figure Start--------------------------------
\begin{figure}
\centerline{
\hfill
\includegraphics[height=0.9\linewidth,angle=270,clip=true]{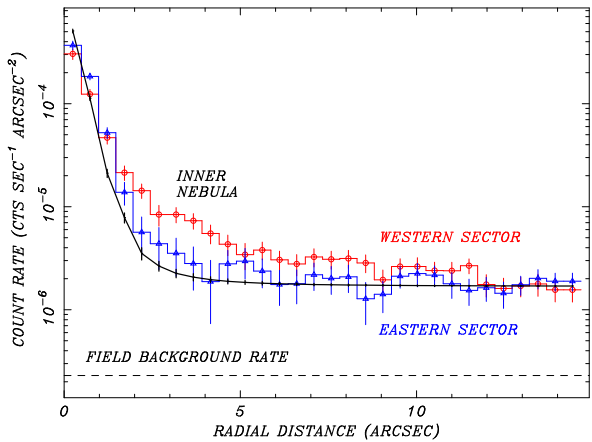}
\hfill
}
\caption{\chandra\ ACIS radial profiles around the candidate pulsar
for the western ({\it red\/}) and eastern ({\it blue}) sectors
compared to the point spread function (PSF; {\it black line}). The PSF
is normalized to the peak of the emission and scaled vertically to
match the data around $15''$ radius to highlight deviation from a pure
point source due to the inner nebula emission, clearly evident in the
western sector (see text).}
\label{fig3}
\end{figure}
%-----------------------------Figure End----------------------------------

To improve on the nominal \chandra\ astrometry, we registered the ACIS
image using nine X-ray sources near \snr\ whose coordinates match with
optical objects in the USNO-B1.0 Catalog (Monet \etal\ 2003 -- typical
uncertainty $0\farcs2$) to $<1^{\prime\prime}$ initially. These
calibrators and other field sources were identified and their
coordinates measured using the CIAO software package source detection
tool {\tt wavdetect}. A fit of the X-ray-optical offsets, weighted by
the positional uncertainties of each source, resulted in a mean
correction of $+0\farcs01$ in R.A.  and $-0\farcs34$ in decl., with a
dispersion of $0\farcs3$ in radius, comparable to the statistical
uncertainties in individual source positions. The final X-ray
positions for the 75 sources that have a signal-to-noise ratio $\geq
3$ are given in Table~1, along with the magnitudes of detected optical
counterparts (see \S2.3). The corrected position of the pulsar
candidate (source \#59) is R.A. = $18^{\rm h}13^{\rm m}35^{\rm
s}\!.166$, decl. = $-17^{\circ}49^{\prime}57^{\prime\prime}\!.48$
(J2000) with a $1\sigma$ error radius of $\approx 0\farcs3$. These
coordinates lie less than $1^{\prime}$ from the maximum probability
centroid of \tev, and well within its $1\sigma$ extent radius of
$2\farcm2$.

The morphology and location of the X-ray source indicates with all but
certainty a young, energetic pulsar/PWN system; this conclusion is
further supported by the following X-ray spectral analysis. We
identify three regions from which to extract spectra: a) the point
source \psr, b) an inner nebula, and c) the greater PWN, which
comprises the bulk of the diffuse emission. In the following we define
each region and tally the extracted counts in the $2-10$~keV energy
band, below which the spectra are highly attenuated due to the large
column density (as determined by the spectral fitting below).

For the point-source spectrum, we extract a total of 864 photons from
a $2\farcs0$ radius aperture centered on the source peak; the diffuse
emission produces a negligible background (19 counts; $\approx 2\%$)
in this region as determined by the counts in a concentric
$2.5^{\prime\prime} \leq r < 15^{\prime\prime}$ annulus. For the inner
nebula we use a $6^{\prime\prime} \times 8^{\prime\prime}$ elliptical
extraction region centered at R.A. = $18^{\rm h}13^{\rm m}34^{\rm
s}\!.89$, decl. = $-17^{\circ}49^{\prime}56^{\prime\prime}\!.9$
(J2000) with a position angle of $240^\circ$. After accounting for the
local background (28\%) in a $1^{\prime}$ diameter circle with the
elliptical region removed, and excising the point source, this yields a
total of 223 photons. For the nebula as a whole, we define an
$80^{\prime\prime}$ radius circle, offset from the point source and
centered at R.A. = $18^{\rm h}13^{\rm m}36^{\rm s}\!.50$, decl. =
$-17^{\circ}49^{\prime}35^{\prime\prime}\!.6$ (J2000), encompassing
the bulk of the nebula extent. With the point source and oval feature
regions excluded, we collect a total of 6638 photons. The background
for the PWN spectrum is extracted from a $56^{\prime\prime}$ radius
aperture offset $2\farcm3$ due north of the source region and
containing 1363 counts.

%\begin{longtable}{lccccrc}
\begin{deluxetable}{lccccrc}
\tablewidth{0pt}
\tabletypesize{\scriptsize}
\tablecaption{Sources in \chandra\ ACIS-I ObsID 6685\label{sources}}
\tablehead{
\colhead{\#} & \colhead{R.A.} & \colhead{Decl.} & \colhead{Counts} &
\colhead{\hfill HR$^a$} & \colhead{$R^b$}  \\
\colhead{} & \colhead{(J2000)} & \colhead{(J2000)} & \colhead{} &
\colhead{} & \colhead{(mag)}  }
\startdata
\s11 & 18 12 50.502 & $-$17 51 13.21 & $25.8 \pm 6.2$ & $+$0.02   & 18.66 \\
\s12 & 18 12 51.980 & $-$17 55 18.71 & $45.0 \pm 8.4$ & $-$0.22   & 16.42 \\
\s13 & 18 12 53.156 & $-$17 56 01.70 & $18.6 \pm 5.7$ & $-$0.45   & 16.63 \\
\s14 & 18 12 53.240 & $-$17 44 45.11 & $24.0 \pm 7.3$ & $+$0.20   & \dots\s4 \\
\s15 & 18 12 53.356 & $-$17 50 22.54 & $44.6 \pm 7.7$ & $-$0.34   & 14.74 \\
\s16 & 18 12 56.116 & $-$17 44 24.61 & $30.6 \pm 7.2$ & $+$0.33   & $>20.5$\s4 \\
\s17 & 18 12 58.332 & $-$17 53 11.61 & $70.1 \pm 8.9$ & $-$0.43   & 15.87 \\
\s18 & 18 12 58.675 & $-$17 46 01.39 & $37.3 \pm 7.3$ & $-$0.22   & 17.40 \\
\s19 & 18 12 59.194 & $-$17 55 25.45 & $162.4 \pm 13.5$ & $-$0.78 & 15.32 \\
10 & 18 13 01.468 & $-$17 49 57.13 & $13.6 \pm 4.2$ & $-$0.13     & 18.72 \\
11 & 18 13 01.859 & $-$17 45 43.95 & $18.6 \pm 5.5$ & $+$0.00     & 16.96 \\
12 & 18 13 03.095 & $-$17 48 45.65 & $19.9 \pm 5.0$ & $-$0.10     & 18.72 \\
13 & 18 13 03.336 & $-$17 47 26.32 & $21.9 \pm 5.2$ & $-$0.14     & \dots\s4 \\
14 & 18 13 08.075 & $-$17 44 31.02 & $16.2 \pm 5.1$ & $-$0.12     & \dots\s4 \\
15 & 18 13 09.351 & $-$17 45 48.31 & $16.3 \pm 4.9$ & $-$0.30     & 18.82 \\
16 & 18 13 11.367 & $-$17 58 46.27 & $13.3 \pm 4.1$ & $+$0.00     & $>20.5$\s4 \\
17 & 18 13 11.400 & $-$17 48 56.13 & $36.9 \pm 6.2$ & $-$0.03     & 20.5:\ts \\
18 & 18 13 11.490 & $-$17 53 37.65 & $25.1 \pm 5.1$ & $-$0.54     & 17.27 \\
19 & 18 13 12.064 & $-$17 51 21.62 & $28.5 \pm 5.5$ & $+$0.40     & $>20.5$\s4 \\
20 & 18 13 12.464 & $-$17 47 11.33 & $20.3 \pm 4.9$ & $-$0.36     & 18.51 \\
21 & 18 13 13.245 & $-$17 51 16.18 & $11.2 \pm 3.5$ & $-$0.50     & 16.70 \\
22 & 18 13 13.427 & $-$17 53 13.06 & $31.6 \pm 5.7$ & $+$0.94     & $>20.5$\s4 \\
23 & 18 13 13.709 & $-$17 44 50.22 & $37.8 \pm 7.1$ & $+$0.72     & $>20.5$\s4 \\
24 & 18 13 14.206 & $-$17 53 43.53 & $272.4 \pm 16.6$ & $+$0.78   & 15.20 \\
25 & 18 13 14.293 & $-$17 45 37.77 & $79.6 \pm 9.6$ & $-$0.22     & 16.06 \\
26 & 18 13 14.443 & $-$17 58 30.62 & $21.7 \pm 5.2$ & $-$0.52     & 17.97 \\
27 & 18 13 14.473 & $-$17 54 02.28 & $13.0 \pm 3.7$ & $-$0.82     & 15.71 \\
28 & 18 13 14.818 & $-$17 49 42.28 & $23.2 \pm 4.9$ & $-$0.79     & 20.0:\ts \\
29 & 18 13 16.446 & $-$17 49 14.06 & $19.2 \pm 4.5$ & $-$0.90     & 15.09 \\
30 & 18 13 17.596 & $-$17 45 57.68 & $12.9 \pm 4.1$ & $+$0.44     & $>23.4$\s4 \\
31 & 18 13 18.902 & $-$17 53 39.04 & $18.3 \pm 4.4$ & $-$0.58     & 21.5:\ts   \\
32 & 18 13 20.117 & $-$17 49 36.55 & $10.5 \pm 3.3$ & $-$0.83     & 17.44 \\
33 & 18 13 21.347 & $-$17 53 31.94 & $10.2 \pm 3.3$ & $+$0.20     & $>23.4$\s4 \\
34 & 18 13 21.433 & $-$17 51 03.54 & $13.4 \pm 3.7$ & $-$0.67     & 11.32: \\
35 & 18 13 21.552 & $-$17 49 22.63 & $10.5 \pm 3.3$ & $-$0.82     & 17.30 \\
36$^c$ & 18 13 21.919 & $-$17 51 36.26 & $9.5 \pm 3.2$ & $-$0.20  & 16.24 \\
37$^c$ & 18 13 22.489 & $-$17 53 50.29 & $34.9 \pm 6.0$ & $+$0.89 & 16.57 \\
38 & 18 13 23.208 & $-$17 49 28.13 & $19.4 \pm 4.5$ & $+$0.68     & 22.3:\ts\\
39 & 18 13 23.313 & $-$17 53 26.33 & $14.1 \pm 3.9$ & $-$0.33     & 13.8:\ts  \\
40 & 18 13 23.595 & $-$17 52 29.50 & $12.6 \pm 3.6$ & $+$0.23      & $>22.5$\s4 \\
41$^c$ & 18 13 23.720 & $-$17 50 40.48 & $214.5 \pm 14.7$ & $+$0.42& 14.94  \\
42$^c$ & 18 13 23.797 & $-$17 53 18.76 & $32.1 \pm 5.7$ & $-$0.52 & 11.85 \\
43$^c$ & 18 13 24.447 & $-$17 52 56.71 & $16.4 \pm 4.1$ & $-$0.53 & 14.68 \\
44 & 18 13 26.022 & $-$17 55 54.13 & $9.7 \pm 3.2$ & $+$0.20       & $>20.5$\s4 \\
45 & 18 13 26.600 & $-$17 51 43.81 & $22.5 \pm 4.8$ & $+$0.65      & 20.2:\ts \\
46 & 18 13 27.176 & $-$17 47 36.05 & $18.0 \pm 4.5$ & $-$0.52     & 16.64 \\
47$^c$ & 18 13 27.501 & $-$17 50 48.79 & $13.2 \pm 3.7$ & $-$0.60 & 17.72 \\
48 & 18 13 28.495 & $-$17 57 35.29 & $20.5 \pm 4.7$ & $-$0.65     & 16.54 \\
49 & 18 13 28.598 & $-$17 48 38.58 & $17.0 \pm 4.2$ & $+$0.78      & $>23.4$\s4 \\
50 & 18 13 29.381 & $-$18 00 16.43 & $22.0 \pm 5.6$ & $+$0.06      & 20.5:\ts \\
51 & 18 13 30.181 & $-$17 51 10.27 & $219.9 \pm 14.9$ & $+$0.99    & $>23.4$\s4 \\
52$^c$ & 18 13 30.543 & $-$17 48 49.60 & $11.1 \pm 3.5$ & $-$0.83 & 16.40 \\
53 & 18 13 31.255 & $-$17 55 01.68 & $12.6 \pm 3.6$ & $-$1.00     & 13.06 \\
54 & 18 13 31.832 & $-$17 50 48.59 & $15.4 \pm 4.1$ & $+$0.76      & $>23.4$\s4 \\
55 & 18 13 32.024 & $-$17 47 49.99 & $10.4 \pm 3.5$ & $+$0.17      & 20.82  \\
56 & 18 13 32.830 & $-$17 56 28.85 & $10.3 \pm 3.3$ & $-$0.33     & 18.41  \\
57 & 18 13 33.301 & $-$17 58 55.53 & $300.0 \pm 18.2$ & $+$0.96    & $>20.0$\s4 \\
58$^c$ & 18 13 33.830 & $-$17 51 48.78 & $13.6 \pm 3.7$ & $-$0.57 & 16.12 \\
59 & 18 13 35.166 & $-$17 49 57.48 & $934.0 \pm 31.0$ & $+$0.99    & $>23.4$\s4 \\
\enddata
\tablecomments{Table continues on the next column.}
%\label{srctable}
\end{deluxetable}

\setcounter{table}{0}

\begin{deluxetable}{lccccrc}
\tablewidth{0pt}
\tabletypesize{\scriptsize}
\tablecaption{--- Continued ---}
\tablehead{
\colhead{\#} & \colhead{R.A.} & \colhead{Decl.} & \colhead{Counts} &
\colhead{\hfill HR$^a$} & \colhead{$R^b$}  \\
\colhead{} & \colhead{(J2000)} & \colhead{(J2000)} & \colhead{} &
\colhead{} & \colhead{(mag)}  }
\startdata
60 & 18 13 35.317 & $-$18 00 13.73 & $29.0 \pm 6.4$ & $-$0.26     & 19.81 \\
61 & 18 13 39.426 & $-$17 54 13.94 & $20.5 \pm 4.6$ & $-$0.65     & 16.1:\ts \\
62$^c$ & 18 13 40.364 & $-$17 51 10.06 & $13.1 \pm 3.7$ & $-$0.47 & 18.02 \\
63 & 18 13 40.473 & $-$17 55 12.87 & $9.5 \pm 3.2$ & $+$0.27       & $>20.5$\s4 \\
64 & 18 13 40.557 & $-$17 46 01.20 & $17.9 \pm 4.8$ & $+$1.00      & $>23.4$\s4 \\
65 & 18 13 41.213 & $-$17 51 15.40 & $58.5 \pm 7.7$ & $+$0.34      & 18.2:\ts \\
66 & 18 13 42.150 & $-$17 45 13.89 & $27.9 \pm 6.2$ & $-$0.06     & 21.24 \\
67 & 18 13 43.315 & $-$17 58 51.43 & $31.7 \pm 6.3$ & $+$0.73      & $>20.5$\s4 \\
68 & 18 13 44.025 & $-$17 49 36.21 & $12.9 \pm 4.0$ & $+$1.00      & $>23.4$\s4 \\
69 & 18 13 46.428 & $-$17 58 34.88 & $263.4 \pm 16.8$ & $-$0.17   & 16.24\\
70 & 18 13 47.460 & $-$17 54 31.06 & $14.7 \pm 4.0$ & $-$0.65     & 15.65 \\
71 & 18 13 47.574 & $-$17 57 00.86 & $71.1 \pm 8.9$ & $-$0.31     & 13.64 \\
72 & 18 13 48.404 & $-$17 52 58.78 & $33.9 \pm 6.0$ & $-$0.72     & 16.18 \\
73 & 18 13 49.121 & $-$17 47 35.97 & $18.1 \pm 4.8$ & $-$0.24     & 17.54 \\
74 & 18 13 50.531 & $-$17 57 01.82 & $12.9 \pm 4.1$ & $-$0.05     & $>20.0$\s4 \\
75 & 18 14 01.415 & $-$17 51 13.13 & $22.7 \pm 5.3$ & $-$0.03     & 16.7:\ts \\
\enddata
\tablecomments{Table includes  background subtracted counts for all point
sources in the ACIS-I field that have a signal-to-noise ratio (S/N) 
greater than 3,
as determined by the CIAO software  package source detection tool {\tt 
wavdetect}.}
\tablenotetext{a}{Hardness ratio defined as HR = $(N_h-N_s)/(N_h+N_s)$, 
where $N_s$ and $N_h$ are the counts measured in the
$0.3-2$~keV and $2-10$~keV energy band, respectively.}
\tablenotetext{b}{Ellipsis means probable optical counterpart exists, but not 
measured due to blending. Colon indicates measurement uncertain due to blending.}
\tablenotetext{c}{USNO-B1.0 catalog counterpart used as an astrometric 
calibrator.}
%\label{srctable}
%\end{longtable}
\end{deluxetable}

Spectra from each region were grouped with a minimum of 15 counts per
spectral channel and fitted using the XSPEC software in the $2-10$~keV
energy band; the results are summarized in Table~2 and Figure~4.  The
PWN emission is well-characterized (but not uniquely) by an absorbed
power-law model with best-fit photon index of $\Gamma = 1.3(1.1-1.6$;
90\% confidence interval), averaged over the nebula, and a hydrogen
column density of $N_{\rm H} \approx 9.8(8.7-11) \times
10^{22}$~cm$^{-2}$, a relatively high value implying substantial
absorption along the line of sight. The spectrum of the putative
pulsar is fitted by a power-law model with the column density fixed to
the value derived from the high-significance PWN spectrum. The
best-fit model yields a surprisingly similar index, $\Gamma = 1.3\pm
0.3$, not unlike that found for the Vela pulsar, and to be expected
for a high-$\dot E$ rotation-powered pulsar (cf. Gotthelf 2003). The
absorbed $2-10$~keV fluxes for the putative pulsar (PSR) and PWN are
$F_{PSR} = 1.3 \times 10^{-12}$~ergs~cm$^{-2}$~s$^{-1}$ and $F_{PWN} =
5.6 \times 10^{-12}$~ergs~cm$^{-2}$~s$^{-1}$, respectively. These
results are consistent with the \asca\ spectral measurements for the
composite spectrum (e.g., Brogan \etal\ 2005). The spectral
distribution of the inner nebula (IN) photons is not well-constrained,
but a fit with the power-law model yields a somewhat flatter index
than that of the PWN ($\Gamma = 0.4^{+0.4}_{-0.7}$), with a flux of
$F_{IN} \sim 4 \times 10^{-13}$~ergs~cm$^{-2}$~s$^{-1}$.

The point-source flux is constant over the 30~ks observation.  The
ACIS frame time of 3.2~s does not permit a search for the signal of a
typical rotation-powered pulsar in this observation. A search for slow
pulsations by means of a fast Fourier transform (FFT) yielded no
evidence of a signal with a $3\sigma$ upper-limit to the pulsed fraction
of $27\%$ for a Nyquist-limited sinusoidal period of $P > 6.5$~s. A
constraint on shorter periods is provided by the \asca\ data. A timing
analysis revealed the source to have a constant flux over the 100~ks
spanned by the 30~ks of net exposure; no periodic variability was
detected in the period range 0.125~s to 1000~s (cf. Brogan et
al. 2005). This is not surprising since the emission is dominated by
the PWN and a pulsed signal would be washed out. Given the relative
fluxes of the various components resolved by \chandra\ but blended by
the \asca\ point source response function, the $3\sigma$ upper limit
on any modulation\footnote{The $3\sigma$ upper-limit of $38\%$
reported in Brogan \etal\ 2005 is corrected here for the PWN flux
contamination in the \asca\ beam.}  is a rather unconstraining 44\%
for a sinusoidal signal for $P>0.125$~s.

\begin{deluxetable}{lccc}
%\rotate
\tablewidth{0pt}
\tabletypesize{\scriptsize}
\tablecaption{\snr\ \chandra\ Spectral Fits\label{spectra}}
\tablehead{
\colhead{Model}    & \colhead{Pulsar}& \colhead{PWN}& \colhead{Inner}\\
\colhead{Parameter}&                 &              & \colhead{Nebula
Only}
}
\startdata
$N_{\rm H}$ ($10^{22}$ cm$^{-2}$) & 9.8(fixed)           & 9.8(8.7,11.0)        & 9.8 (fixed)\\ 
$\Gamma$                          & 1.3(1.0,1.6)         & 1.3(1.1,1.6)         & 0.4(-0.3,0.8)\\
PL Flux\tablenotemark{a}          & $1.3\times 10^{-12}$ & $5.6\times 10^{-12}$ & $4\times 10^{-13}$\\
$\chi2$(DoF)                      &  43(38)              & 122(129)             & 15(17)\\
\enddata
\tablenotetext{a}{Absorbed flux in the 2--10 keV band in units of ergs
cm$^{-2}$ s$^{-1}$.}
\tablecomments{Uncertainties are 90\% confidence for two interesting
parameters.}
\end{deluxetable}

%-----------------------------Figure Start--------------------------------
\begin{figure}
\centerline{
\includegraphics[height=0.9\linewidth,angle=270,clip=true]{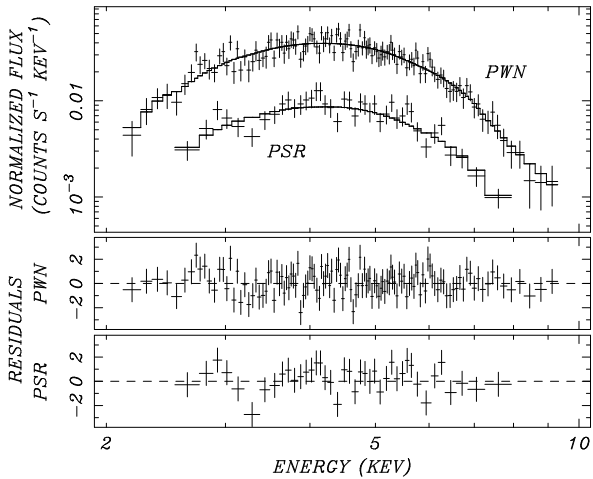}
}
\caption{Chandra ACIS spectrum of \psr\ and its nebula, both fitted to
an absorbed power-law model. Residuals from the best-fit model
(summarized in Table~2) for each data set are shown in the bottom
panels.}
\label{fig4}
\end{figure}
%-----------------------------Figure End----------------------------------

\subsection{Search for a Radio Pulsar}

As part of a project to search for pulsar counterparts to a number of
interesting radio sources described in Helfand \etal\ (2006), on 2005
September~8 we observed \tev\ with the ATNF Parkes telescope in NSW,
Australia.  Our pointing position, $\mbox{R.A.}=18^{\rm h}13^{\rm
m}19\fs4$, $\mbox{decl.}=-17\arcdeg 54'46''$, was $6\farcm1$ away from
the precise position of the subsequently identified \chandra\ source.
We used for this search the central beam of the Parkes multibeam
receiver operating at a central frequency of 1374~MHz, with 96
frequency channels spanning a total bandwidth of 288~MHz in each of
two polarizations (see, e.g., Manchester \etal\ 2001).  During the
integration time of 30~ks, total-power samples were recorded every
0.25~ms to magnetic tape for off-line analysis.

We analyzed the data with standard pulsar searching techniques
implemented in the PRESTO software package (Ransom 2001; Ransom \etal\
2002).  We searched the dispersion measure range
0--3344\,cm$^{-3}$\,pc (twice the maximum Galactic DM predicted for
this line-of-sight in the Cordes \& Lazio 2002 electron density
model), while maintaining close to optimal time resolution (scattering
of the radio pulses due to multipath propagation is of no consequence
here, being about 1~ms for the expected $\mbox{DM} \sim
300$~cm$^{-3}$~pc at the distance of 4.5~kpc which is justified in \S3.1).  We
first excised the worst of the radio frequency interference in the
data, and searched for pulsars having a large range of duty cycles and
spin periods between 0.5~ms and $\sim 5$~s.  We also searched for
pulsars whose spin period could have changed moderately during the
observation (due either to very large intrinsic spin-down or a binary
companion).  The search followed very closely that described in more
detail in Camilo \etal\ (2006).  We did not identify any promising
pulsar signal in this search.

Because of the $6\farcm1$ offset between our pointing position and the
location of the X-ray pulsar candidate (\S~2.1), the sensitivity of
our search was $\approx0.7$ that of an on-source search (the
full-width at half-maximum of the telescope beam is $14\farcm4$).
Using the standard modification to the radiometer equation for a
conservative duty cycle for the pulsations of 10\%, and accounting for
a sky temperature at this location of 15~K, we were sensitive to
long-period pulsars ($P \ga 50$\,ms) having a period-averaged flux
density at 1.4~GHz of $S_{1400} >0.07$~mJy, with the limit becoming
progressively worse for shorter pulse periods.  For a distance of
$\sim 4.5$~kpc, this corresponds to a pseudo-luminosity limit of
$L_{1400} \equiv S_{1400} d^2 \la 1.4$~mJy~kpc$^2$.  This is nearly a
factor of 3 above the detected $L_{1400}$ for the young pulsar in 3C58
(Camilo \etal\ 2002a), which has the smallest known luminosity among
young pulsars, but is comparable to or below that of other young,
low-luminosity pulsars (see, e.g., Camilo \etal\ 2002b).

\subsection{Optical Observations}

The field containing \snr\ was observed on 2005 July 5 using the 2.4m
Hiltner Telescope of the MDM Observatory on Kitt Peak, Arizona.  A
thinned, backside illuminated SITe CCD with a spatial scale of
$0^{\prime\prime}\!.275$ per $24 \mu m$ pixel was used with an
$R$-filter to cover a $9^{\prime}\!.4$ field centered on \snr\ in
seeing of $1^{\prime\prime}\!.5$.  In Figure~5 we show a portion of
the combined 18 minute exposure centered on \psr.  Taken under
photometric conditions, the image was calibrated using Landolt (1992)
standard stars.  An astrometric solution for the image was derived in
the reference frame of the USNO-B1.0 catalog (Monet \etal\ 2003) using
44 stars that have an rms dispersion of $0^{\prime\prime}\!.45$ about
the fit.  Since this is the same reference frame to which the X-ray
coordinates were corrected, we can say confidently that there is no
optical counterpart of \psr\ to a $3\sigma$ limiting magnitude of
$R=23.4$.  Given the large column density to the source, the
equivalent of $\sim 20$ mag of visual extinction, this limit is
unconstraining.  No counterpart is apparent in the 2MASS near-IR or
Spitzer GLIMPSE (Benjamin et al. 2003) mid-IR images either.

%-----------------------------Figure Start--------------------------------
\begin{figure}
\centerline{
\hfill
\includegraphics[width=0.8\linewidth,angle=0,clip=true]{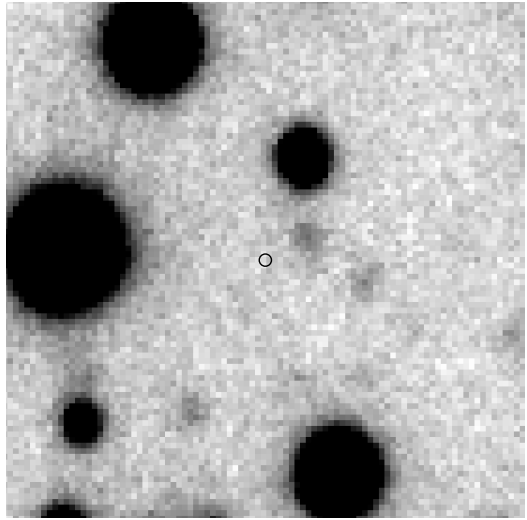}
\hfill
}
\caption{A portion of the
$R$-band CCD image centered at the location of \psr, obtained
with the 2.4m Hiltner Telescope on 2005 July 5.
The seeing is $1^{\prime\prime}\!.5$, and the $3\sigma$
limiting magnitude is $R=23.4$.  The $1\sigma$
error circle of radius $0^{\prime\prime}\!.3$
marks the location of the putative X-ray pulsar,
R.A. = $18^{\rm h}13^{\rm m}35^{\rm s}\!.166$,
decl. = $-17^{\circ}49^{\prime}57^{\prime\prime}\!.48$ (J2000).
North is up and east is to the left in this
$25^{\prime\prime} \times 25^{\prime\prime}$ display.
}
\label{fig5}
\end{figure}
%-------------

%-----------------------------Figure Start--------------------------------
\begin{figure}
\centerline{
\hfill
\includegraphics[width=0.9\linewidth,angle=0,clip=true]{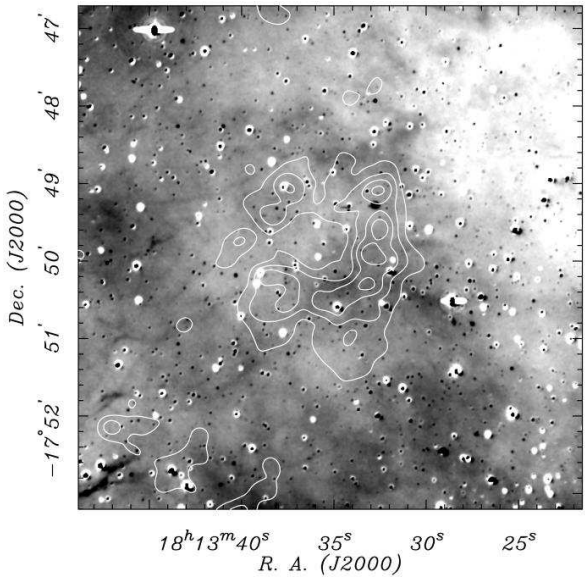}
\hfill
}
\caption{H$\alpha$ image of the field of \snr, obtained
with the 2.4m Hiltner Telescope on 2005 July 5, with
the VLA radio contours overlaid. Point-like artifacts from 
imprecisely subtracted and saturated stars remain.
}
\label{fig6}
\end{figure}
%-------------

We also obtained 72 minutes of exposure through a rest-frame H$\alpha$
filter with a bandpass of 100 \AA.  A scaled version of the $R$-band
image was subtracted from the H$\alpha$ image to highlight the diffuse
emission by removing stars to the extent possible. Figure~6 shows
the resulting H$\alpha$ image, with the VLA radio contours of \snr\
superposed.  While diffuse H$\alpha$ emission is present throughout
the image, there is no indication from its morphology that any of it
is associated with \snr.  Again, given the extinction to the source,
this null result is unsurprising.

We searched for likely optical counterparts to the other X-ray sources
given in Table~1.  $R$ magnitudes were obtained from the USNO-B1.0
catalog or, for fainter or blended stars, from our own CCD image,
which covers only part of the X-ray field. In this dense stellar
field, it is possible that some proposed counterparts are chance
coincidences, especially near the edge of the field where X-ray
positions are less accurate.  Bright sources \#51 and \#69 are evidently
flare stars as seen from their X-ray light curves, although \#51 does
not have an optical counterpart to the very deep limit $R>23.4$.
Sources without optical counterparts generally have hardness ratio
$>+0.2$, consistent with being AGN or other distant objects.

\section{Discussion}

Our investigation of \snr\ with \chandra\ makes a compelling case that
the X-ray emission from this remnant is dominated by the energy output
of a young, rotation-powered pulsar. In this section we explore the
energetics of this source and constraints on its properties that can
be developed in the absence of a detection of pulsed emission. We also
discuss the possible implications for the origin of TeV emission from
other examples of supernova products.

\subsection{ The Distance to \snr }

The X-ray absorbing column density we derive here for both the
putative pulsar and its PWN, $N_{\rm H}=9 \times 10^{22}$ cm$^{-2}$,
is consistent with that found earlier from the \asca\ data (Brogan et
al 2005; Ubertini et al 2005). Brogan et al. provide a thorough review
of the data relevant to the source distance, including the molecular
column density of $N({\rm H}_2) = 8 \times 10^{22}$ cm$^{-2}$ to a
distance of $\sim 4$~kpc in this direction, and the total \ion{H}{1}
column through the Galaxy at this longitude of $2 \times 10^{22}$
cm$^{-2}$. Given that a distance of $4-5$~kpc is roughly half the way
to the edge of the \ion{H}{1} disk in this direction, the total X-ray
absorption is easily explained if \snr\ lies at or just beyond
4~kpc. The source's non-detection at low radio frequencies is
consistent with it lying in the vicinity of, or beyond, the giant star
formation region W33 which resides $18^{\prime}$ to the south at an
estimated distance of 4.3~kpc.  Given these data and the plausibility
of finding a young, core-collapse remnant in a star formation region,
we adopt a distance of 4.5~kpc.

\subsection{The X-ray Energetics}

That the putative pulsar is highly energetic is indicated by the large
nebula-to-pulsar flux ratio of $F_{PWN}/F_{PSR} = 4.3$ in the
$2-10$~keV energy band; only rotation-powered pulsars with spin-down
energy loss rates above $\dot E \approx 4 \times
10^{36}$~ergs~s$^{-1}$ have a ratio this large or greater (Gotthelf
2003). Furthermore, the total 2--10~keV luminosity of $L_x = 1.74
\times 10^{34} \ d^2_{4.5}$ ergs s$^{-1}$ from \snr\ corresponds to
$\dot E \sim 10^{37}$~ergs s$^{-1}$ according to the correlation of
Possenti \etal\ (2002),\ placing this object among the Galaxy's ten
most energetic pulsars. In particular, \snr\ bears a striking
similarity to SNR~G106.6+2.9, a radio shell undetected in X-rays that
contains the energetic pulsar PSR~J2229+6114 with $\dot E = 1.8 \times
10^{37}$~ergs~s$^{-1}$ and flux ratio $F_{PWN}/F_{PSR} = 9$ (Halpern
\etal\ 2001).

\subsection{The origin of TeV Gamma-rays}

The gamma-ray luminosity of \tev\ at energies above 200 GeV is $4.4
\times 10^{34} \ d^2_{4.5}$~ergs s$^{-1}$, nearly identical to its
$2-10$~keV X-ray luminosity. This is the highest $L_{TeV}/L_x$ ratio
for any of the confirmed PWN gamma-ray emitters; the ratios are 0.5
for PSR~1509--58, roughly 0.3 for SNR~G0.9+0.1 and Vela X, and 0.06
for the Crab Nebula.  Upper limits to the TeV flux from two other
prominent young PWNe in the composite remnants SNR~G11.2--0.3 and
SNR~G29.7--0.3 (Kes~75) can be derived from the HESS Galactic Plane
survey (Aharonian \etal\ 2006); they also imply $L_{TeV}/L_x \simlt
1$.  As Funk \etal\ (2006) show, the TeV emission from \snr\ can be
fitted simultaneously to the hard X-ray data from INTEGRAL (Ubertini
\etal\ 2005) and the $0.5-10$ keV X-ray spectrum\footnote{Note that
the X-ray fluxes in their Figure 6 should be reduced by about 20\% to
account for the fraction of the emission originating in the pulsar
itself.} with an inverse-Compton model, although a low-energy break in
the electron spectrum is required to explain the absence of a radio
counterpart to the PWN.

We regard the proximity of W33 as a likely explanation for the
relatively high TeV flux from \snr. We find it noteworthy that a
number of HESS sources are found where high-energy particle
accelerators are located near \ion{H}{2} regions, while other such
accelerators without proximate sources of low-energy photons are weak
or undetected at TeV energies. For example, the Crab Nebula has the
lowest $L_{TeV}/L_x$ for a PWN observed to date and lies more than
$5^\circ$ from the Galactic plane; likewise, Kes~75, an isolated SNR
housing a very young and energetic pulsar but with no nearby
\ion{H}{2} regions, has $L_{TeV}/L_x < 0.2$ and is undetected by
HESS. On the other hand, SNR~G0.9$-$0.1 near the Galactic Center, PSR
1509$-$58 near the \ion{H}{2} region G320.5$-$1.4, and \snr\ near W33
are all easily detected.

In addition, HESS~J1640$-$465 is a new PWN candidate associated with
the shell-type SNR G338.3$-$0.02 and is adjacent to a bright
\ion{H}{2} region (Funk \etal\ 2007).  Those authors quote a
$\Sigma-D$ distance for this remnant of 8~kpc, and therefore suggest
that it is unassociated with the \ion{H}{2} region, whose kinematic
distance places it at 3~kpc. However, the $\Sigma-D$ relation is
notoriously unreliable (Green 2005) and the 8~kpc distance would imply
that HESS J1640$-$465 is considerably more luminous than the Crab
Nebula, whereas a 3~kpc distance would imply $L_{TeV} \sim 2.5 \times
10^{34}$~ergs s$^{-1}$, similar to the other PWN detected to date. In
addition, the source HESS~J1834$-$087, coincident with the shell-type
SNR~G23.3$-$0.3, also lies within $9^{\prime}$ of a bright \ion{H}{2}
region, while the historical remnant SN1006, the first SNR from which
hard X-ray synchrotron flux was detected, has not been seen at TeV
energies despite an intensive search (Aharonian \etal\ 2005); it lies
over $14^\circ$ from the Galactic plane. This accumulating anecdotal
evidence suggests that a systematic evaluation of the local optical/IR
flux in the vicinity of known particle accelerators might point the
way to detecting more TeV sources. {\it GLAST\/} will be invaluable
for distinguishing between leptonic and hadronic TeV emissions
mechanisms (e.g., Funk 2007); the evidence adduced here suggests the
inverse Compton mechanism may be preferred for the majority of HESS
detections.

\acknowledgements Support for this work was provided by the National
Aeronautics and Space Administration through \chandra\ Grants
SAO~GO6-7052X (D.J.H.) and G06-7057X (E.V.G.) issued by the \chandra\
X-ray Observatory Center, which is operated by the Smithsonian
Astrophysical Observatory for and on behalf of the National
Aeronautics Space Administration under contract NAS8-03060.

\end{document}